\documentclass{aa}

\usepackage{txfonts}
\usepackage[dvips]{graphicx}
\usepackage{natbib}
\bibpunct{(}{)}{;}{a}{}{,} 

\begin{document}

\title{The small dispersion of the mid IR -- hard X-ray correlation in AGN
       \thanks{Based on ESO observing programme 075.B-0844(C)}}
\author{Hannes Horst \inst{1,2}
  \and Alain Smette \inst{1}
  \and Poshak Gandhi \inst{3,1}
  \and Wolfgang J. Duschl \inst{2,4}}
\authorrunning{H. Horst et al.}
\offprints{H. Horst, \email{hhorst@eso.org}}
\institute{European Southern Observatory, Casilla 19001, Santiago 19, Chile
  \and Zentrum f\"ur Astronomie, ITA, Universit\"at Heidelberg, 
       Albert-Ueberle-Str. 2, 69120 Heidelberg, Germany
  \and Institute of Astronomy, Madingley Road, Cambridge CB3 0HA, UK
  \and Steward Observatory, The University of Arizona, 933 N. Cherry Ave, 
       Tucson, AZ 85721, USA}
\date{Received 00.00.0000 / Accepted 00.00.0000}
\abstract{\emph{Context}. We investigate mid-infrared and X-ray properties of the dusty torus 
          in unification scenarios for active galactic nuclei.

         \emph{Aims}. We use the relation between mid IR and hard X-ray luminosities to 
          constrain AGN unification scenarios.

         \emph{Methods}. With VISIR at the VLT, we have obtained the currently highest angular
           resolution ($0 \farcs 35$ FWHM) narrow-band mid infrared images of
           the nuclei of 8 nearby Seyfert galaxies. Combining these 
           observations with X-ray data from the literature we study the 
           correlation between their mid IR and hard X-ray luminosities.

         \emph{Results}. We find that the rest frame $12.3 \, \mu$m ($L_{\mathrm{MIR}}$) 
          and 2-10 keV ($L_{\mathrm{X}}$) luminosities are correlated at a 
          highly significant level. The best fit power-law to our data is
          $\log{L_{\mathrm{MIR}}} \propto \left(1.60 \pm 0.22 \right) 
          \log{L_{\mathrm{X}}}$, showing a much smaller dispersion than earlier
          studies.

         \emph{Conclusions}. The similarity in the $\log{L_{\mathrm{MIR}}}$ / 
          $\log{L_{\mathrm{X}}}$ ratio between Sy1s and Sy2s even using high 
          angular resolution MIR data implies that the similarity is intrinsic 
          to AGN and not caused by contamination from extra-nuclear emission. 
          This supports clumpy torus models. The exponent of the correlation 
          constrains the inner geometry of the torus.
          
}

\maketitle

\section{Introduction}

The unification model for active galactic nuclei (AGN) interprets the 
different appearance of Seyfert 1 (Sy1) and Seyfert 2 (Sy2) galaxies uniquely 
as the result of an orientation effect \citep{anton93,barthel94,urry96}. The 
central engine is considered to be surrounded by an optically and geometrically
thick molecular torus. Associated with this torus are large masses of dust 
that supposedly reprocess the X-ray and UV emission from the accretion disk 
and re-emit it in the mid infrared (MIR) regime \citep{pier93}. 

It is thus very attractive to search for correlations between IR continuum and 
hard X-ray emission in order to test the unification scenarios for AGN. A tight
correlation between the $10.5 \, \mu$m continuum and the absorption-corrected 
2-10 keV luminosities for 8 nearby Seyfert galaxies was reported by 
\citet{krabbe01} using $1\farcs 2$ resolution MIR data. More recently 
\citet{lutz04} found a correlation between the rest frame $6 \, \mu$m 
luminosity and the absorption-corrected hard X-ray luminosity for a sample 
of $71$ AGN. This sample was comprised of objects for which $24 \arcsec$ 
angular resolution ISOPHOT spectra and hard X-ray observations were available; 
in particular it does not contain Compton-thick objects. However, the authors
reported two problems that the unification model faces when compared  with 
their observations. (I) The scatter of the relation is about an order of 
magnitude larger than expected from the results of \citet{krabbe01}. (II) 
There is no significant difference between early type (Sy types 1 to 1.5) and 
late type (Sy types 1.8 to 2) objects in the average ratio of mid-infrared to 
hard X-ray emission, as would be expected from an optically and geometrically 
thick torus dominating the mid-IR AGN continuum. 

Here we study this relation through rest wavelength $12.3 \, \mu$m photometry 
of AGN, improving upon \citet{krabbe01} and \citet{lutz04} by a
factor of 3 and 80, respectively, in terms of angular resolution. The 
wavelength was chosen as, among the bands accessible from the ground, it 
best traces hot and warm dust.

Throughout this paper we assume $H_{0} = 72\,$km/s, $\Omega_{\Lambda} = 0.73$
and $\Omega_{\mathrm{m}} = 0.27$ \citep{spergel03}.

\section{Observations and data analysis} \label{data}

Between March and July 2005 we observed 8 nearby Seyfert nuclei (see Table
\ref{sample}), selected from the sample used by \citet{lutz04}, with VISIR 
\citep{lagage04} at the VLT. The individual objects were selected as to avoid a
strong bias toward higher luminosities with higher redshift. All observations 
were carried out under good photometric and optical seeing conditions 
($\leq 0 \farcs 8$) and with  airmass below 1.3. For all observations we used 
the parallel chop/nod mode of VISIR with a chopping throw of $8 \arcsec$, thus 
receiving all three beams on the detector. We observed each object in three 
different narrow-band filters in order to obtain the rough shape of its MIR 
spectral energy distribution (SED). These filters were chosen as to avoid 
emission and absorption lines, in particular the $12.8 \, \mu$m NeII line and 
the broad $9.7 \, \mu$m silicate feature which in cases of very strong 
absorption can extend to $\lambda \approx 12.5 \, \mu$m \citep{dorschner78}. 
\object{NGC 7314} was only observed in two different filters. Total integration
times ranged between 1.2 and 9.2 min per filter and per object.

We reduced science and standard star frames using the pipeline written by Eric
Pantin (private communication). To eliminate glitches, the pipeline applies a 
bad pixel mask and removes detector stripes. Subsequently we removed background
variations using a 2 dimensional polynomial fit. Here we treated each nodding 
cycle separately as the background pattern sometimes changes between two 
consecutive cycles. The count rate for one full exposure was calculated as the 
mean of all 3 beams from all nodding cycles of this exposure. As an error 
estimate we use the standard deviation of these. In order to minimise the 
effect of residual sky background we chose relatively small apertures 
($\approx 10$ pixels $= 1 \farcs 27$) for the photometry and corrected the obtained 
count rates using the radial profiles of standard stars. Finally we calibrated 
our photometry using the same standard stars. These were observed within 1.5 
hrs of the science exposures and with a maximum airmass difference of 0.15. Our
observational results are shown in Table \ref{obs}. Comparing the FWHM of 
standard star and science exposures yields no indication that our sources are 
extended. Thus our observations indeed probe the physical scales given in Table
\ref{sample}.
\begin{table}
   \caption{Basic parameters of our sample of galaxies. Heliocentric redshifts 
           $z$ were taken from the NED, Seyfert types are according to 
           \citet{veron00}. The absorption corrected hard X-ray luminosities 
           we compiled from the literature. Where multiple 
           observations were available the average was taken. In these cases 
           the reported uncertainties correspond to 1/2 of the peak-to-peak
           variabilities. References are $^{\textrm{a}}$\citet{turner89}, 
           $^{\textrm{b}}$\citet{gondoin01}, $^{\textrm{c}}$\citet{lutz04}, 
           $^{\textrm{d}}$\citet{reeves04}, $^{\textrm{e}}$\citet{terashima02},
           $^{\textrm{f}}$\citet{ho01}, $^{\textrm{g}}$\citet{reynolds04}, 
           $^{\textrm{h}}$\citet{steenbrugge03}, 
           $^{\textrm{i}}$\citet{cappi05}, $^{\textrm{k}}$\citet{risaliti02},
           $^{\textrm{l}}$\citet{landi01}, $^{\textrm{m}}$\citet{kraemer04}, 
           $^{\textrm{n}}$\citet{risaliti00}, $^{\textrm{o}}$\citet{smith96}.
           In the case of \object{PG 2130+099} only one measurement could be 
           found. Here we assume a variability of a factor of 2. The last 
           column contains the physical scales resolved by VISIR, computed for 
           an angular resolution of $0".35$, which is a typical value for our 
           observations (see table \ref{obs}). 
           $^*$For a discussion of the classification of NGC 4579 see 
           section \ref{results}.}
   \begin{center}
    \begin{tabular}{lcclc}
      \hline\hline
      Object & $z$ & Sy & log $L_{2-10 \, \textrm{\tiny{keV}}}$ & 
      Scale [pc]\\
      \hline
      \object{Fairall 9} & 0.047 & 1.2 & 
      $43.84 \pm 0.15^{\textrm{\tiny{a,b,c}}}$ & 320 \\
      \object{NGC 526a} & 0.019 & 1.9 &
      $43.14 \pm 0.35^{\textrm{\tiny{a,c,k,l}}}$ & 135 \\
      \object{NGC 3783} & 0.010 & 1.5 & 
      $43.12 \pm 0.15^{\textrm{\tiny{a,c,d}}}$ & 70 \\
      \object{NGC 4579} & 0.005 & 1.9$^{*}$ &
      $41.18 \pm 0.22^{\textrm{\tiny{c,e,f,i}}}$ & 35 \\
      \object{NGC 4593} & 0.009 & 1.0 & 
      $42.89 \pm 0.15^{\textrm{\tiny{a,c,g,h}}}$ & 65 \\
      \object{PKS 2048-57} & 0.011 & 2.0 & 
      $43.07 \pm 0.15^{\textrm{\tiny{c,o}}}$ & 80 \\
      \object{PG 2130+099} & 0.062 & 1.5 & 
      $43.68 \pm 0.15^{\textrm{\tiny{c,n}}}$ & 415\\
      \object{NGC 7314} & 0.005 & 1.9 & 
      $42.19 \pm 0.15^{\textrm{\tiny{a,c,k,m}}}$ & 35 \\
      \hline
    \end{tabular}
  \end{center}
  \label{sample}
\end{table}
\begin{table*}
  \caption{Basic observational parameters for all data used: Object name, 
           observing date (in 2005), filter, flux and full width half maximum
           for  standard star (STD) and science observations. As error 
           estimates we use the standard deviation of all contributing beams 
           (see section \ref{data} for details). The FWHM are the average of 
           all beams of one exposure. All science and corresponding STD 
           exposures were obtained within a maximum airmass difference of 0.15.
           $^*$The central wavelengths for the individual filters are 
           $9.82 \, \mu$m for SIVref1, $10.49 \, \mu$m for SIV, $11.25 \, \mu$m
           for PAH, $11.88 \, \mu$m for PAH2ref2, $12.27 \, \mu$m for NeIIref1 
           and $12.81 \, \mu$m for NeII. Their transmission curves are shown in
           the VISIR User's manual at http://www.eso.org/instruments/visir/.}
  \begin{center}
    \begin{tabular}{lclccc|lclccc}
      \hline\hline
      Object & Obs. Date & Filter$^*$ & Flux [mJy] & 
      \multicolumn{2}{c|}{FWHM [$\arcsec$]}&
      Object & Obs. Date & Filter$^*$ & Flux [mJy] & 
      \multicolumn{2}{c}{FWHM [$\arcsec$]} 
      \\
      & (MM-DD)&  & & STD & Obj & & (MM-DD) & & & STD & Obj \\
      \hline
      \object{Fairall 9} & 07-21 & SIV & $256 \pm 5$ & 0.32 & 0.35 &
      \object{NGC 4593} & 04-30 & SIV & $331 \pm 29$ & 0.30 & 0.31 \\
      \object{Fairall 9} & 07-21 & NeIIref1 & $330 \pm 18$ & 0.38 & 0.37 &
      \object{NGC 4593} & 04-30 & PAH2ref2 & $335 \pm 26$ & 0.33 & 0.33 \\
      \object{Fairall 9} & 07-21 & NeII & $306 \pm 10$ & 0.40 & 0.37 & 
      \object{NGC 4593} & 04-30 & NeIIref1 & $382 \pm 73$ & 0.34 & 0.36 \\
      \object{Fairall 9} & 07-21 & SIV & $235 \pm 13$ & 0.32 & 0.34 &
      \object{PKS 2048-57} & 06-10 & PAH2ref2 & $883 \pm 53$ & 0.38 & 0.40 
      \\
      \object{NGC 526a} & 07-20 & SIV & $199 \pm 26$ & 0.41 & 0.30 &
      \object{PKS 2048-57} & 06-10 & NeIIref1 & $1040 \pm 64$ & 0.41 & 0.43
      \\
      \object{NGC 526a} & 07-20 & NeIIref1 & $275 \pm 55$ & 0.41 & 0.35 &
      \object{PKS 2048-57} & 06-10 & SIV & $655 \pm 28$ & 0.37 & 0.39 \\
      \object{NGC 526a} & 07-20 & SIV & $199 \pm 22$ & 0.41 & 0.32 &
      \object{PG 2130+099} & 06-10 & PAH2 & $152 \pm 8$ & 0.39 & 0.38 \\
      \object{NGC 3783} & 04-17 & SIV & $568 \pm 46$ & 0.29 & 0.31 &
      \object{PG 2130+099} & 06-10 & SIVref1 & $115 \pm 20$ & 0.35 & 0.36 
      \\
      \object{NGC 3783} & 04-17 & PAH2ref2 & $632 \pm 22$ & 0.34 & 0.35 &
      \object{PG 2130+099} & 06-10 & NeII & $179 \pm 31$ & 0.41 & 0.43 \\
      \object{NGC 3783} & 04-17 & NeIIref1 & $722 \pm 67$ & 0.34 & 0.36 &
      \object{PG 2130+099} & 06-10 & PAH2 & $174 \pm 17$ & 0.39 & 0.40 \\
      \object{NGC 4579} & 04-30 & SIV & $67 \pm 17$ & 0.30 & 0.25 &
      \object{NGC 7314} & 07-20 & SIV & $75 \pm 29$ & 0.40 & 0.36 \\
      \object{NGC 4579} & 04-30 & PAH2ref2 & $69 \pm 14$ & 0.33 & 0.29 &
      \object{NGC 7314} & 07-20 & PAH2 & $75 \pm 22$ & 0.36 & 0.37 \\
      \object{NGC 4579} & 04-30 & NeIIref1 & $61 \pm 21$ &  0.34 & 0.30 &
      \object{NGC 7314} & 07-20 & SIV & $87 \pm 37$ & 0.40 & 0.34 \\
      \object{NGC 4579} & 04-30 & SIV & $64 \pm 12$ & 0.30 & 0.32 & & & &
      & & \\ \hline
    \end{tabular}
  \end{center}
  \label{obs}
\end{table*}

\section{Results} \label{results}

\begin{figure*}
  \begin{center}
    \includegraphics[width=0.7\textwidth,height=3.7cm,clip=]{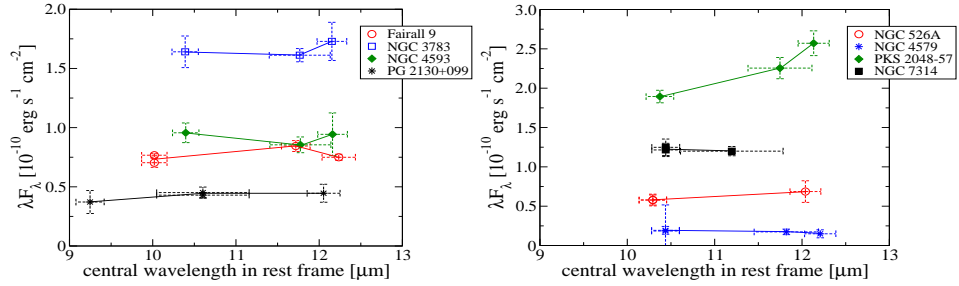}
  \end{center}  
  \caption{The MIR SEDs of our sample of galaxies. Early- and late-type 
             Seyferts are shown in the left- and right-hand panel respectively.
             Horizontal error bars show the band width of the used filters.
             The flux densities of \object{NGC 7314} have been offset by 
             $+1 \cdot 10^{-10}$erg s$^{-1}$ cm$^{-2}$ for clarity.}
  \label{seds}
\end{figure*}

\begin{figure*}
  \begin{center}
     \includegraphics[width=0.7\textwidth,height=2.7cm,clip=]{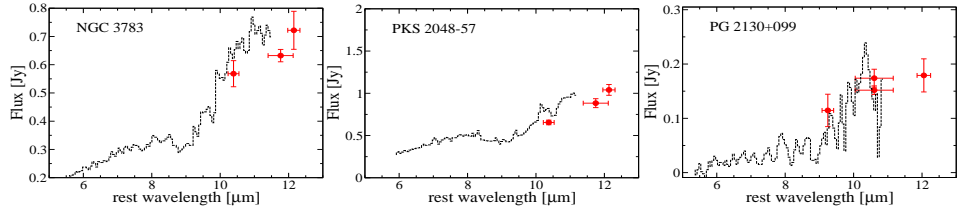}
  \end{center}
   \caption{Comparison of ISOPHOT-S spectra (dashed histograms) taken from 
             \citet{rigo99} and our VISIR photometry (filled circles). 
             Horizontal error bars show the band width of the filters used.}
  \label{rigo}
\end{figure*}

Figure \ref{seds} shows all photometric data points taken during our observing 
programme. For comparison Fig. \ref{rigo} shows our measurements together 
with ISOPHOT spectra taken from \citet{rigo99} for these three sources common 
to both samples. Among our data the SED of \object{PKS 2048-57} is the only one
displaying a strong $9.7 \, \mu$m silicate feature which, however, does not
significantly affect our measurement of the continuum flux in the NeIIref1 
filter. This is important as it validates our result on the MIR -- hard X-ray 
correlation (see below).

One of the objects in our sample -- \object{NGC 4579} -- has peculiar MIR 
properties. Comparing the ISOPHOT $6 \, \mu$m flux reported in \citet{lutz04} 
of $S_{\mathrm{6 \mu m}} = 96\,$mJy to our VISIR flux levels (see Table 
\ref{obs}) leads to the conclusion that either this object has a peculiar MIR 
SED that does not rise toward longer wavelengths as expected for typical
AGN or that the bulk of IR emission  ($\gtrsim 70\%$) does not originate in 
the central point source. In both cases it is highly questionable whether the 
latter is dominated by a molecular torus. Moreover \citet{contini04} reports 
the MIR emission of \object{NGC 4579} to be dominated by thermal emission of
shock-heated ISM in the vicinity of the nuclear region rather than by 
reprocessed X-ray and UV emission from the central engine itself. These 
peculiarities plus the fact that \object{NGC 4579} has also been classified as 
a low-ionisation nuclear emission region (LINER) by \citet{keel83} and
\citet{filippenko85} led us to exclude it from the MIR -- hard X-ray 
correlation investigated here. \object{NGC 7314}, the other low luminosity 
object in our sample, harbours a regular Seyfert nucleus \citep{lumsden04}.

For all objects in our sample we calculated the $12.3 \, \mu$m rest frame 
luminosity by using the filter with the central wavelength closest to 
$12.3 \, \mu$m and assuming a flat SED in this spectral region. Then we 
compiled absorption-corrected 2-10 keV measurements for our sample from the 
literature (see Table \ref{sample} for details). When several such measurements
were available we use the mean luminosity for the MIR -- hard X-ray 
correlation and take half of the peak-to-peak-variability as a conservative 
estimate of the uncertainty in the hard X-ray luminosity. Only one such 
measurement could be found for \object{PG 2130+099} for which we therefore 
assumed a variability of a factor of 2. As we derive the luminosity distances 
of all targets solely from their redshifts we assume an uncertainty of 3 Mpc 
due to possible deviations from the Hubble flow. This is applied to the error 
bars of both X-ray and MIR luminosities. 

\begin{figure}
  \begin{center}
    \includegraphics[width=0.3\textwidth,clip]{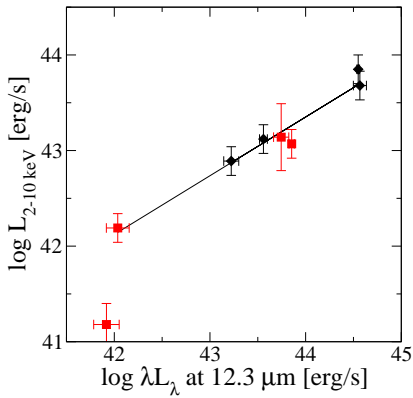}
  \end{center}
  \caption{Correlation of MIR to absorption-corrected hard X-ray luminosities 
     for our VISIR sample. Filled diamonds are early-type Seyferts (Type Sy 
     1.5 or smaller) while filled squares are late-type Seyferts. The power law
     displayed is the best fit obtained when omitting \object{NGC 4579} 
     (see section \ref{results} for details).}
  \label{correl}
\end{figure}

The resulting MIR -- hard X-ray correlation we obtain for our VISIR sample is 
shown in Fig. \ref{correl}. Pearson's coefficient is $r = 0.98$ meaning that
the correlation is highly significant. The best fit power-law to our data is 
$\log{L_{\mathrm{MIR}}} = \left( -25.4 \pm 9.5 \right) + \left(1.60 \pm 0.22 
\right) \cdot \log{L_{\mathrm{X}}}$ with measurement errors considered in the
fit. Including \object{NGC 4579} into the fit yields an exponent of 
$\sim 1.28$ rather than 1.6 and $r = 0.95$.

\section{Discussion and conclusions} \label{discussion}

We have presented new high-resolution MIR observations of 8 nearby Seyferts, 
obtained with VISIR at the VLT. We found that their rest-frame $12.3 \,\mu$m  
and absorption-corrected 2-10 keV luminosities are strongly correlated 
following $L_{\mathrm{MIR}} \propto L_{\mathrm{X}}^{1.6}$. Early- and late-type
Seyfert galaxies follow the same relation. At the low-luminosity end the 
correlation is not well defined as one of the two object's MIR radiation is 
very probably not dominated by a dusty torus. 

Both \citet{krabbe01} and \citet{lutz04} found a correlation close to 
$L_{\mathrm{MIR}} \propto L_{\mathrm{X}}^1$. The exponent different to
the one we find is most probably due to the difference in angular resolution 
($24 \arcsec$ for ISOPHOT, $1 \farcs 2$ for Krabbe et al.'s sample and 
$ \sim 0 \farcs 35$ for VISIR) and to the choice of the rest frame wavelength 
as the intervening medium is expected to be more optically thin at 
$12.3 \, \mu$m than at shorter wavelengths. In particular, using the 
corresponding ISOPHOT $6 \; \mu$m data for our sample also yields an exponent 
close to 1. In addition we find a dispersion similar to the one in 
\citet{krabbe01}; the larger one reported by \citet{lutz04} can be attributed 
to their relatively poor angular resolution and the fact that the emission at 
$6 \, \mu$m is slightly more dependent on inclination effects 
\citep[e.g. see][]{hoenig06}. On the other hand we confirm the absence of a 
significant difference in the $L_{\mathrm{MIR}}$ / $L_{\mathrm{X}}$ ratio 
between early and late-type objects found by both other studies. At our high 
angular resolution this is clearly in disagreement with smooth torus models, in
which a flux difference of one order of magnitude is expected for type 1 and 
type 2 AGN of the same bolometric luminosity \citep[e.g.][]{pier93}. This 
cannot be accounted for by dust clumps situated in the ionisation cone 
\citep{efstat95} because their contribution to the nuclear MIR flux should not 
exceed 50\%  \citep[see][]{galliano05}. Clumpy torus models 
\citep[e.g.][]{nenkova02} do not  predict such an offset and seem to be 
compatible with our results.

\citet{hoenig06} recently used 3D radiative transfer modelling to demonstrate 
that clumpy tori can appear as optically thin in the MIR with most of the 
radiation originating in the innermost part of the torus. In their model 
individual clouds are optically thick but their filling factor is very small. 
Consequently lines-of-sight toward the innermost clouds are usually not blocked
by other clouds. This implies that $L_{\mathrm{MIR}}$ is proportional to the 
surface area $A$ of the inner edge of the torus and the covering factor of 
clouds projected on $A$. Our finding that Sy1s and Sy2s follow the same 
$L_{\mathrm{X}}$ -- $L_{\mathrm{MIR}}$ relation is a strong indication that the
torus appears optically thin at $12.3 \, \mu$m. 

The correlation we find then requires a dependency of $A$ or the covering 
factor on $L_{\mathrm{X}}$ and thus the bolometric luminosity 
$L_{\mathrm{Bol}}$. The simplest explanation for this observed effect is 
flaring: If the torus' inner edge is defined by the dust sublimation radius 
$r_0$, we obtain $A = 2 \pi r_0 \cdot H(r_0)$ where $H(r)$ is the scale height 
of the cloud distribution. For a flaring disk $H(r)$ can be written as 
$H(r) \propto r^{\alpha}$ with $\alpha$ being constant.  This assumption yields
$A \propto r_0^{1+\alpha}$. The dust sublimation radius $r_0$ follows 
$r_0 \propto L_{\mathrm{Bol}}^{0.5} \propto L_{\mathrm{X}}^{0.5}$. This leads 
to $L_{\mathrm{MIR}} \propto L_{\mathrm{X}}^{0.5 + 0.5\alpha}$. Our 
observations yield $L_{MIR} \propto L_{X}^{1.6}$ and thus $\alpha \approx 2$. 
Including \object{NGC 4579} results in $\alpha \approx 1.5$.

Another possibility is that its scale height depends on the radial accretion 
rate $\dot{M}$. The torus model presented by \citet{beckert04} predicts 
$H(r)/r \propto \dot{M}^{\, 0.5}$. For AGN with comparable radiative 
efficiencies $\dot{M}$ is proportional to $L_{\mathrm{X}}$. This yields 
$L_{\mathrm{MIR}} \propto A \propto L_{\mathrm{X}}^{1.5}$ which is very close 
to the correlation we find.

The high angular resolution of our MIR observations indicates that the 
similarity of the $L_{\mathrm{MIR}}$ / $L_{\mathrm{X}}$ ratio for Sy1s and Sy2s
is intrinsic rather than caused by extra-nuclear emission. Although 
contributions of dust clumps situated in the ionisation cone cannot be ruled 
out they should not dominate the nuclear MIR flux. Assuming our sample is 
representative of the general AGN population our results thus support recent 
models of clumpy dust tori \citep[e.g.][]{nenkova02,hoenig06} and, 
furthermore, indicate that the filling factor of clouds is small. The geometry 
of the torus can be constrained by the slope of the 
$L_{\mathrm{X}}$ -- $L_{\mathrm{MIR}}$ correlation. Clumpy tori with low 
filling factors might also account for the similarity of \emph{Spitzer} spectra
of type 1 and type 2 QSOs \citep{sturm06}.

\begin{acknowledgements}
We thank Thomas Beckert, Sebastian H\"onig and Eric Pantin for 
helpful discussions and an anonymous referee for comments and suggestions.
This research made use of the NASA/IPAC Extragalactic Database (NED) which 
is operated by the Jet Propulsion Laboratory, California Institute of 
Technology, under contract with the National Aeronautics and Space 
Administration.
\end{acknowledgements}

\bibliographystyle{aa}
\bibliography{mybiblio}

\end{document}